\documentclass[twocolumn,showpacs,preprintnumbers,amsmath,amssymb]{revtex4}

\usepackage{graphicx}
\usepackage{dcolumn}
\usepackage{bm}

\begin{document}


\title{The Latent Heat of Single Flavor Color
Superconductivity in a Magnetic Field}

\author{{Ping-ping Wu $^{\rm a}$}, Hang He$^{\rm
a}$, Defu Hou $^{\rm a}$, Hai-cang Ren $^{\rm a,b}$}

\affiliation{{$^{\rm a}$Institute of Particle Physics, Huazhong
Normal University,Wuhan 430079, China}\\{$^{\rm b}$Physics
Department, The Rockefeller University,1230 York Avenue, New York,
NY 10021-6399}
}%

\begin{abstract}
We calculate the energy release associated with first-order phase
transition between different types of single flavor color
superconductivity in a magnetic field.
\end{abstract}

\maketitle

At sufficiently high baryon density, a hadronic matter will be
squeezed into a degenerate system of quarks, which becomes a color
superconductor(CSC) at sufficiently low temperature\cite{MAKT}.

The most plausible place in the universe where the color
superconductivity may be implemented is the interior of compact
stars.They have masses about $0.5M_{\bigodot}-2.7 M_{\bigodot}$
and are believed to have radii of order 10km. Near the surface
their density is around normal nuclear density and raises to
several times higher in the core region. There a quark matter of
chemical potential $\mu\sim400-500MeV$ may form.

The color superconductivity can influence the observable signature
of a compact star in several ways. Firstly, it will impact on the
quark matter equation of state. Secondly, the energy gap, as well as
the Goldstone and/or Higgs modes associated to the long range interaction
the transport properties of the star. Thirdly, the presence of
strong magnetic field renders the superconducting phase transition
first order. The latent heat release at the transition may lead to
observable energy burst which will be discussed in this report.

While the color-flavor-locked(CFL) phase, which pairs different
quark flavors, is the ground state at ultra high baryon density
(chiral limit), the situation is far more complicated at the
moderate density in a compact star. The mass of s quarks and the
electrical neutrality constraint create a substantial mismatch of
the Fermi momenta among different flavors which in turn reduces
the available phase space of Cooper pairing. The single-flavor CSC
(pairing within each flavor), however, is free from such a
limitation. The smallness of the single-flavor gaps because of the
nodal lines of the pairing force, makes the single-flavor phases
relevant for late age stars, when the temperature is sufficiently
low. Among the four canonical phases, the spherical
color-spin-lock(CSL) and nonspherical planar, polar and
A\cite{SWR,T,equality1}, CSL is the most favored one in the
absence of a magnetic field.

The presence of a magnetic field offsets the energy balance among
the four canonical single flavor pairings. Only the spherical CSL
phase has electromagnetic Meissner effect\cite{SWR} and
nonspherical phases: polar,A and planar phases only shield part of
the magnetic field. So if a quark matter cools down through the
critical temperature of the single flavor paring in a magnetic
filed, forming CSL state will cost extra work to exclude magnetic
fluxes from the bulk. Therefore, the magnetic contribution to the
free energy may favor the non-spherical states. In a previous
work\cite{magne}, we have explored the consequences of the absence
of the electromagnetic Meissner effect in a nonspherical CSC phase
of single-flavor pairing and have obtained the phase diagram with
respect to the temperature and the magnetic field. Computing the
latent heat released acrossing the phase boundaries is the main
subject of the present work. We shall employ the same
approximation in\cite{magne} by ignoring the masses of pairing
quarks.

We work with a NJL-like effective action whose Lagrangian density
reads\cite{Alford}:

\begin{eqnarray}
\label{njllike}
{\cal L}=\bar\psi(-\gamma_\nu\partial_\nu+\mu\gamma_4)\psi+G\bar\psi\gamma_\nu
T^l\psi\bar\psi\gamma_\nu T^l\psi
\end{eqnarray}
with $T^l=\frac{1}{2}\lambda^l$, $G>0$ an effective coupling and
introduce the condensate:
\begin{equation}
\Phi=<\bar\psi_C\Gamma^c\lambda^c\psi> \label{diquark}
\end{equation}
where $\psi$ is the quark field, $\psi_C=\gamma_2\psi^*$ is its
charge conjugate, $\lambda^c$ with $c=2,5,7$ are antisymmetric
Gell-Mann matrices and $\Gamma^c$ is a $4\times 4$ spinor matrix.
We may choose $\Gamma^5=\Gamma^7=0$ for the polar and A phases,
$\Gamma^2=0$ for the planar phase but none of $\Gamma^c$'s
vanishes for CSL phase. We find the pressure of each flavor under
mean field approximation:

\begin{eqnarray}
\nonumber P=&-&\frac{2}{\Omega}\sum_{{\bf k}}(k-\mu-E_{\bf k})
-\frac{1}{\Omega}\sum_{\bf
k}(k-\mu-|k-\mu|)\\\nonumber&+&\frac{2T}{\Omega} \sum_{\bf
k}\ln(1+\exp(-\frac{|k-\mu|}{T}))-\frac{9}{4G}\Delta^2\\&+&\frac{4T}{\Omega}
\sum_{\bf k}\ln\left(1+\exp\left(-\frac{E_{\bf
k}}{T}\right)\right), \label{pressure}
\end{eqnarray}
where $E_{\bf k}=\sqrt{(k-\mu)^2+\Delta^2f^2(\theta)}$ with
$\theta$ the angle between the momentum ${\bf k}$ and a prefixed
spatial direction and $\Delta$ the gap parameter. The function
$f(\theta)$ is given by
\begin{equation}
f(\theta)=\left\{\begin{array}{ll}
\begin{gathered}
1,
\hspace{0.2cm}\hbox{for CSL phase}\\
\end{gathered}
\hfill\\
\begin{gathered}
\sqrt{\frac{3}{4}(1+\cos^2\theta)},
\hspace{0.2cm}\hbox{for planar phase}\\
\end{gathered}
\hfill\\
\begin{gathered}
\sqrt{\frac{3}{2}}\sin\theta,
\hspace{0.2cm}\hbox{for polar phase}\\
\end{gathered}
\hfill\\
\begin{gathered}
\sqrt{3}\cos^2\frac{\theta}{2}.
\hspace{0.2cm}\hbox{for A phase}\\
\end{gathered}
\end{array}
\right. \label{function}
\end{equation}
Maximizing the pressure with respect to $\Delta$, we obtain the
gap equation $\left(\frac{\partial
P}{\partial\Delta^2}\right)_\mu=0$, which determines the
temperature dependence of the gap, $\Delta(T)$. Substituting
$\Delta(T)$ back to (\ref{pressure}), we find that $P_n<P_{\rm
A}<P_{\rm polar}<P_{\rm planar}<P_{\rm CSL}$ up to the transition
temperature $T_c$.

The diquark condensate (\ref{diquark}) for CSL breaks the gauge
symmetry $SU(3)_c\times U(1)_{\rm em}$ completely. A nonspherical
condensate, however, breaks the gauge symmetry partially and the
Meissner effect is incomplete. Among the residual gauge group,
which is the gauge transformation that leaves the diquark operator
inside (\ref{diquark}) invariant,there exists a U(1)
transformation, $\psi\to
e^{-\frac{i}{2}\lambda_8\theta-iq\phi}\psi$ with $q$ the electric
charge of $\psi$, $\theta=-2\sqrt{3}q\phi$ for the polar and A
phases and $\theta=4\sqrt{3}q\phi$ for the planar phase. The
corresponding gauge field, ${\cal A}_\mu$ is identified with the
electromagnetic field in the condensate and is related to the
electromagnetic field $A$ and the 8-th component of the color
field $A^8$ in the normal phase through a $U(1)$ rotation
\begin{eqnarray}
{\cal A}_\mu &=& A_\mu\cos\gamma-A_\mu^8\sin\gamma\nonumber\\
{\cal V}_\mu &=&
A_\mu\sin\gamma+A_\mu^8\cos\gamma\label{transform}
\end{eqnarray}
where the mixing angle $\gamma$ is given by
$\tan\gamma_{\rm{polar,A}}=2\sqrt{3}q(e/g)$ and
$\tan\gamma_{\rm{planar}}=4\sqrt{3}q(e/g)$ with $g$ the QCD
running coupling constant. The 2nd component of (\ref{transform})
${\cal V}=0$ because of the Meissner effect and thereby imposes a
constraint inside a nonspherical CSC, $A_\mu^8=-A_\mu\tan\gamma$,
which implies that:
\begin{equation}
{\bf B}^8=-{\bf B}\tan\gamma \label{constraint}
\end{equation}

The thermal equilibrium in a magnetic field $H\hat{\bf z}$ is
determined by minimizing the Gibbs free energy density,
\begin{equation}
{\cal G}=-P+\frac{1}{2}B^2+\frac{1}{2}\sum_{l=1}^8(B^l)^2-BH
\label{gibbs}
\end{equation}
with respect to $\Delta$, $B$ and $B^l$. Ignoring the induced magnetization of quarks,
the pressure $P$ is given by (\ref{pressure}), with $\Delta$ given by the solution
of the gap equation. For a nonspherical CSC pairing, the minimization with respect to $B$
and $B^l$ is subject to the constraint (\ref{constraint}).
For a hypothetical quark matter of one flavor
only, we find that
\begin{equation}
{\cal G}_{\rm min.,j}=-P_j-\frac{1}{2}\eta_jH^2
\end{equation}
with $j=n$, CSL, polar, A and planar, where $\eta_n=1$, $\eta_{\rm
CSL}=0$ and $\eta_j=\cos^2\gamma_j$ for a nonspherical CSC. The
phase corresponding to minimum among ${\cal G}_{\rm min}$'s above
wins the competition and transition from one phase to another is
first order below $T_c$.

For a quark matter of different flavors with Cooper pairing within
each flavor, the Gibbs free energy density can be written as
\begin{equation}
{\cal G}=-P-\frac{1}{2}\eta H^2.
\end{equation}
where $P$ is the total pressure of all flavors. We have $\eta=1$
if all flavors are normal and $\eta=0$ if all flavors are CSL. If
one flavor is in a nonsherical state and others are normal,
$\eta=\cos^2\gamma$ with $\gamma$ the mixing angle of the
nonspherical phase. If more than one flavors are in nonspherical
phases, $\eta=\cos^2\gamma$ with $\gamma$ their common mixing
angle. If their mixing angle were different, we would end up with
$B=B^8=0$, in order to compromise the constraints
(\ref{constraint}) of all nonspherical states involved, making
them less favored than CSL. It was shown in \cite{magne} that only
four phases, I-IV(shown in Table.I and Fig.1), need to be
considered in both two and three flavor quark matters. There is a
first order phase transition from one of them to another for
$0<T<T_c$.
\begin{table*}
\caption{The possible phases involving spin-one CSC of a quark matter with
two flavors or three flavors in a magnetic field. \label{tab:table1}}
\begin{ruledtabular}
\begin{tabular}{ccccc}
&I&II&III&IV\\
\hline
 2 flavor&$\rm CSL_u, CSL_d$&$\rm (polar)_u, (planar)_d$&$\rm (normal)_u, (polar)_d$&$\rm (normal)_u, (normal)_d$ \\
 3 flavor&$\rm CSL_u, CSL_{d,s}$&$\rm (polar)_u, (planar)_{d,s}$&$\rm (normal)_u, (polar)_{d,s}$&$\rm (normal)_u, (normal)_{d,s}$\\
\end{tabular}
\end{ruledtabular}
\end{table*}

\begin{figure}[!htb]
\centering
\includegraphics[scale=1, clip=true]{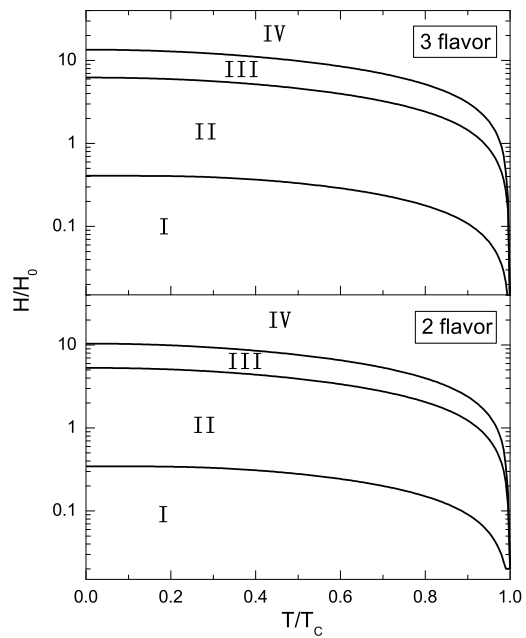}
\caption{The H-T phase diagram for two flavors and three
flavors, where the reference magnetic field $H_0=\mu\Delta_0/\pi$ with $\Delta_0$
the CSL gap at $T=0$.\label{fig:1}}
\end{figure}
The border between two phases are determined by the equation:
\begin{equation}
P_\alpha+\eta_\alpha\frac{H^2}{2}=P_\beta+\eta_\beta\frac{H^2}{2}
\end{equation}
with the subscripts $\alpha$ and $\beta$ labeling the four phases I-IV and the
density of the latent heat released from the phase $\alpha$ to the phase $\beta$ reads
\begin{equation}
Q_{\alpha\beta}=T[S_\alpha(T)-S_\beta(T)],
\label{latent}
\end{equation}
The entropy density $S(T)$ is given by
\begin{eqnarray}
\nonumber S&=&\left(\frac{\partial P}{\partial
T}\right)_{\mu}=\left(\frac{\partial P}{\partial
T}\right)_{\Delta,\mu} +\left(\frac{\partial
P}{\partial\Delta}\right)_{T,\mu}
\left(\frac{\partial\Delta}{\partial T}\right)_{\mu}
=\left(\frac{\partial P}{\partial T}\right)_{\Delta,\mu}
\end{eqnarray}
where the gap equation is employed in the last step.

The temperature dependence of the thermodynamic quantities, $P$
and $S$ can be written in a parametric form. In terms of the
parameter $t=\frac{\Delta(T)}{T}$, the gap equation takes the form
$\ln\frac{\Delta(0)}{\Delta(T)}=h(t)$ with
\begin{eqnarray}
h(t) &=& \int_0^{\pi}d\theta\sin\theta
f^2(\theta)\\&\int_0^{\infty}&dx\frac{1}{\sqrt{x^2+t^2f^2(\theta)}[e^{\sqrt{x^2+t^2f^2(\theta)}}+1]}\nonumber
\end{eqnarray}
It follows that
\begin{equation}
T=\frac{\Delta(0)}{t}e^{-h(t)}. \label{para1}
\end{equation}
Introducing $P_s-P_n\equiv\rho(t)\frac{\mu^2\Delta_0^2}{2\pi^2}$
and $
S_s-S_n=\frac{\mu^2\Delta_0^2}{2\pi^2}\left(\frac{d\rho}{dt}/\frac{dT}{dt}\right)$
with $s$ lableling different pairing states and
$\Delta_0\equiv\Delta_{\rm CSL}(0)$, we have
\begin{equation}
\rho(t)=e^{-2h(t)}[1+2h(t)+4\frac{g(t)}{t^2}-\frac{2\pi^2}{3t^2}]
\end{equation}
with
\begin{equation}
g(t)=\int_0^{\pi}d\theta\sin\theta\int_0^{\infty}dx\ln[e^{-\sqrt{x^2+t^2f^2(\theta)}}+1]
\end{equation}
and the curves $P(T)$ and $S(T)$ may be plotted parametrically
without solving the gap equation for $T>0$. As we did in
\cite{magne}.

Numerically, we identify $\Delta_{0}$ with that of the one-gluon
exchange\cite{T,ogluon} with $\mu=500MeV$ and
$\alpha_s=\frac{g^2}{4\pi}=1$ as a calibration of the parameters in (\ref{njllike}),
We find $\Delta_0\simeq 0.238MeV$, $H_0\simeq 5.44\times 10^{14}$G  for two
flavors, and $\Delta_0\simeq 0.0864MeV$, $H_0\simeq 1.97\times 10^{14}$G for three
flavors. As is shown in Fig.1, the nonspherical phases occupy a significant
portion of the H-T phase diagram for a magnitude of the magnetic
field of order $10^{15}G$. This magnitude of the magnetic field is plausible
in a compact star.

The temperature dependence of the entropy density differences
between all single flavor CSC phases and that of the normal phase
under the same chemical potential $\mu=500MeV$ is shown in Fig.2.
The differences vanish at $T=0$ since all entropies are zero. They
also vanish at $T=T_c$ since the transition is second order there.
 The entropy density of a CSC phase is lower than that of the normal phase
because of the long range order,so the differences are all
negative for $0<T<T_c$. The latent heat densities across the phase
boundaries in the Fig.1 are displayed in the Fig.3, where
adjustment of the chemical potential of each flavor is made to
fulfill the requirement of the charge neutrality. The latent heat
gets a peak value in about 0.67$T_c$-0.72$T_c$ for all curves.

\begin{figure}
\centering
\includegraphics[scale=0.7, clip=true]{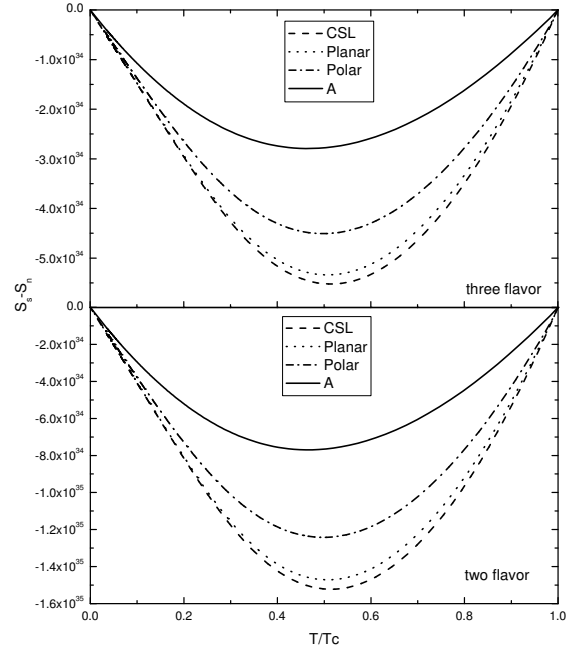}
\caption{The entropy density differences between single flavor CSC
phases and normal phase as a function of the temperature.The unit
is $erg/K/km^3$.\label{fig:entropy1}}
\end{figure}

\begin{figure}
\centering
\includegraphics[scale=0.7, clip=true]{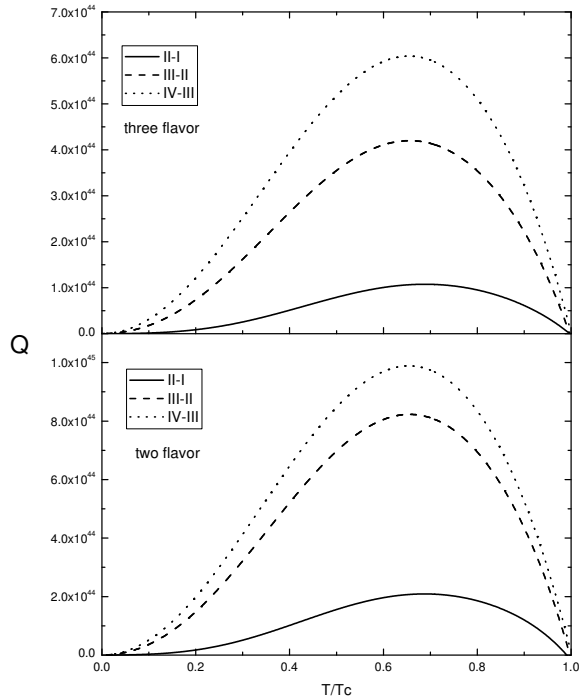}
\caption{The latent heat density dependence of temperature.We know
that $S_s=S_n$ at $T_c$ and
$Q_{\alpha\beta}=T[S_\alpha(T)-S_\beta(T)]$, so latent heat also
vanish at $T_0$ and $T_c$. The unit of Q is
$erg/km^3$.\label{fig:heat1}}
\end{figure}
In the natural unit we are using, the entropy is dimensionless and
the unit of latent heat can be $MeV$, the unit of latent heat
density is $MeV^4=2.088\times10^{41}erg/km^3$. The units have been
transformed to International System of Units in Fig.2 and Fig.3.
For the biggest latent heat density we calculated in three flavor
phase, from IV to III, this value is about
$6.04\times10^{44}erg/km^3$. Then for a compact star with the
radii of quark matter equal to $R(km)$, the energy release is
about $2.53\times10^{45}R^3 erg$. Since $R<10km$ typically, the
latent heats released as the star cools through the phase
boundaries we showed  are smaller than the typical energy release
of $\gamma$ ray burst of order  $10^{51}$ to
$10^{54}erg$\cite{gamma,gamma2}. But they may contribute to weaker
energy bursts such as the X-ray radiation at the later stage of a
compact star\cite{neutron}.

Finally, we would like to comment on the approximations employed.
The two flavor case is certainly unrealistic since it
assumes a large mass of $s$ quarks, $m_s>>\mu$. For three flavors,
our approximation requires that $m_s$ to be sufficiently smaller
than $\mu$. On the other hand $m_s$ cannot be too small in order
for the single flavor CSC to compete with multi-flavor pairings
under the Fermi momentum mismatch, which requires $m_s$ to be of
the same order of magnitude as $\sqrt{\mu\Delta(2SC)}$ or greater,
with $\Delta(2SC)$ the 2SC gap at zero temperature without
mismatch. The phase structure of 2SC in a magnetic field has been
studied in\cite{2sc}. Since $\Delta(2SC)$ is about 5-10 times lower
than $\mu$ for the quark matter we considered, the approximation, though
marginal, may not be too crude as is
suggested by our recent analysis with a realistic $m_s$, 150MeV,
in the single flavor pairing dynamics\cite{mass}. For the highest magnetic field
in the phase diagram of Fig.1, $H\simeq 5\times 10^{15}$G, which implies
$eH/(\mu^2)\simeq 6\times 10^{-5}$. Therefore it is legitimate to ignore
the impact of the magnetic field on the pairing dynamics, unlike the situation
considered in \cite{manuel}.

Throughout the paper we have assumed that the whole volume of the quark matter core of a compact star
undergoes a phase transitions at the same time.
In reality the chemical potential, and consequently the gap and the critical temperature,
change with the distance from the center and the transition at the center and that at the edge
of the core may not be simultaneously. For a typical compact star, the density at the center
is at most two times larger than that on the edge\cite{Glend}
and the quark matter cools very rapidly and reaches isothermal condition in a few hours
because of the high thermal conductivity\cite{lattimer}.
As a crude estimation, we associate the chemical potential of 500MeV to the center of the
quark matter core and define its edge at the radius where the transition temperature drops
by half. Then time scale for the transition at the center to that at the
edge is about a few minutes to a few days depending on the direct Urca process\cite{lattimer}.
This is much shorter than the typical age of a compact star ($10^4-10^6 y $),
justifying our picture of the sudden release of the latent heat. The subject
deserves further study to clarify more quantitatively the interplay between the phase transition
and the structure and the evolution of a compact star.

\begin{acknowledgments}
We would like to extend our gratitude to Bo Feng, Li Wen, Qiu-han
Wang and Xiao-ping Zheng for helpful discussions.The work of D. F.
H. and H. C. R. is supported in part by NSFC under grant Nos.
10575043, 10735040.
\end{acknowledgments}


\begin{thebibliography}{99}

\bibitem{MAKT}M. G. Alford, A. Schmitt, K, Rajagopal and T. Sch$\ddot{a}$fer, Rev. Mod. Phys.
80:1455-1515, 2008  and the references therein.

\bibitem{SWR}A. Schmitt, Q. Wang and D. H. Rischke, Phys. Rev. Lett. {\bf 91}, 242301 (2003).

\bibitem{T}T. Sch$\ddot{a}$fer, Phys. Rev. {\bf D62}, 094007
(2000).
\bibitem{equality1}Andreas Schmitt, Phys. Rev. D{\bf 71}, 054016 (2005).

\bibitem{magne}Bo Feng,De-fu Hou, Hai-cang Ren,and Ping-ping Wu , Phys. Rev. Lett. {\bf 105}, 042001 (2010).

\bibitem{Alford}M. Alford and G. Cowan, J. Phys. G: {\bf 32},
511-528 (2006).

\bibitem{ogluon}D. T. Son, Phys. Rev. D{\bf59}, 094019 (1999); T. Sch$\ddot{a}$fer and F.
Wilczek, Phys. Rev. D{\bf 60}, 114033 (1999); R. D. Pisarski and
D. H. Rischke, Phys. Rev. D{\bf 61}, 074017 (2000); W. E. Brown,
J. T. Liu and H-C Ren, Phys. Rev. D{\bf 62}, 054016 (2000).

\bibitem{gamma}Z. Berezhiani, I. Bombaci, A. Drago, F. Frontera, A. Lavagno,
Astrophys. J. {\bf 586} 1250 (2003).
\bibitem{gamma2}K. S. Cheng, Z.G. Dai, Phys. Rev. Lett. {\bf 77}, 1210 (1996).
\bibitem{neutron}M. Alford, D. Blaschke,A. Drago,T. Klahn, G. Pagliara and J. Schaffner-Bielich,
Nature {\bf 445} E7 (2007).
\bibitem{2sc} Sh. Fayazbakhsh, N. Sadooghi, Phys. Rev. D{\bf 83}, 025026
(2011).

\bibitem{mass}Ping-ping Wu, Defu Hou and Hai-cang Ren, in
preparation.

\bibitem{manuel}E. Ferrer, V. Incera and C. Manuel, Phys. Rev. Lett. {\bf 95}, 152002 (2005);
J. Noronha and I. A. Shovkovy, Phys. Rev. D{\bf 76}, 105030 (2007);
K. Fukushima and H. Warringa, Phys. Rev. Lett. {\bf 100}, 032007 (2008).

\bibitem{Glend} N. K. Glendenning,  {\it Compact Stars } (Springer Press), 1997

\bibitem{lattimer} J. M.  Lattimer, K. A. V. Riper,  Madappa Prakash, and  Manju Prakash, Astrophys. Jour. {\bf 425}:
802(1994)

\end{thebibliography}
\end{document}